\documentclass{aastex}
\usepackage{emulateapj5}
\submitted{Accepted by ApJ, February 2008} 

\def\kms{km~s$^{-1}$}

\def\degree{$^{\circ}$}

\def\ga{\mathrel{\hbox{\rlap{\hbox{\lower4pt\hbox{$\sim$}}}\hbox{$>$}}}}
\def\la{\mathrel{\hbox{\rlap{\hbox{\lower4pt\hbox{$\sim$}}}\hbox{$<$}}}}

\shorttitle{The many streams of the Magellanic Stream}

\shortauthors{S.\ Stanimirovi\'{c} et al.}
\begin{document}

\title{The many streams of the Magellanic Stream}

\author{Sne\v{z}ana Stanimirovi\'{c}\altaffilmark{1}, Samantha
  Hoffman\altaffilmark{1}, Carl Heiles\altaffilmark{2}, 
Kevin A. Douglas\altaffilmark{3}, 
  Mary Putman\altaffilmark{4}, Joshua E. G. Peek\altaffilmark{2}}
\altaffiltext{1}{Department of Astronomy, University of Wisconsin, Madison, WI
  53706; sstanimi\@astro.wisc.edu, shoffman3\@wisc.edu}
\altaffiltext{2}{Department of Astronomy, UC Berkeley, 601 Campbell Hall,
Berkeley, CA 94720; goldston\@astro.berkeley.edu}
\altaffiltext{3}{Space Sciences Laboratory, University of California,
  Berkeley, CA 94720; douglas\@ssl.berkeley.edu}
\altaffiltext{4}{University of Michigan, Department of Astronomy, 500 Church
St., Ann Arbor, MI 48109; mputman\@umich.edu}

\begin{abstract}
We present results from neutral hydrogen (HI) observations of the
tip of the Magellanic Stream (MS), obtained with the Arecibo telescope as a part
of the on-going survey by the Consortium for Galactic studies with the Arecibo
L-band Feed Array.
We find four large-scale, coherent HI streams, extending
continously over a length of 20 degrees, each stream possessing  
different morphology and velocity gradients. 
The newly discovered streams provide  strong support for the
tidal model of the MS formation by \cite{Connors06},
which suggested a spatial and kinematic bifurcation of the MS. 
The observed morphology and kinematics suggest that three of these streams could
be interpreted as a 3-way splitting of the main MS filament, while the fourth
stream appears much younger and may have originated from the Magellanic Bridge.

We find an extensive population of HI clouds at the tip of the MS. 
Two thirds of clouds have an angular size in the range 3.5$'$--10$'$.
We interpret this as being due to thermal instability,
which would affect a warm tail of gas trailing through the Galactic halo
over a characteristic timescale of a few Myrs to a few hundred Myrs.
We show that thermal fragments can survive in the hot halo for a long time,
especially if surrounded by a $<10^{6}$ K halo gas.
If the observed clumpy structure is mainly due to thermal instability, then
the tip of the MS is at a distance of $\sim70$ kpc.
A significant fraction of HI clouds at the tip of the MS 
show multi-phase velocity profiles, indicating the co-existence 
of cooler and warmer gas. 
 \end{abstract}

\keywords{ISM: clouds --- ISM: structure --- radio lines: ISM}

\section{Introduction}

The Magellanic Stream (MS), a 10\degree~wide
tail of neutral hydrogen (HI) emanating from the Magellanic
Clouds and trailing for almost 100\degree~on the sky (Dec $\sim -60$ to
$+20$\degree) \citep{Wannier72,Mathewson74}, is the only clear example of 
a gaseous halo stream in 
the Milky Way's close proximity. While it is well accepted that the MS is the
result of interactions between the Milky Way (MW) and the Magellanic Clouds, the
relative importance of tidal stripping and various kinds of gasdynamical
interactions is still very much under debate.
Most recently several new attempts were made to model the observed HI column density
 and velocity distribution as being due to purely tidal stripping
\citep{Connors06}, or gravitational $+$ hydrodynamical interactions
\citep{Mastropietro05}.
These models focus on reproducing general features in the Magellanic
System, and gradients in HI column density and velocity 
along the MS. 
To add more excitement to this topic, recent estimates of the proper motion
of the Small Magellanic Cloud (SMC) by \cite{Kallivayalil06} and 
the most recent calculations of the Magellanic Cloud's orbits
by \cite{Besla07} suggest, contrary to all previous studies, that the Clouds are only
on their first passage around the MW. 
The new orbits imply that neither tidal nor ram pressure stripping would 
have had
enough time to produce the MS, calling for
alternative formation mechanisms.

The distance to the MS, especially to its tip, or the region
the farthest away from the Magellanic Clouds, is another contentious question and 
varies greatly between models. Under the ram pressure hypothesis the tip has
fallen the farthest toward the MW  and is at a distance of only 25 kpc.
Early tidal models \citep{Gardiner96,Yoshizawa99}
place the tip at a distance of 60-70 kpc. 
The latest tidal simulations
\citep{Connors06} find an even more distant component extending
from 170 to 200 kpc.   
The latest orbit calculations \citep{Besla07}
would also imply a significantly large distance to the MS, $\sim150$ kpc, although it
is not clear where exactly the MS is relative to the Clouds in this framework. 

For many years the MS was viewed as a complex of six discrete
concentrations (labeled as MS I to VI). 
New HI Parkes surveys by \cite{Putman03}
and \cite{Bruns05}, with an angular resolution of 15.5$'$, revealed a more complex nature 
of the MS gas, with a fascinating network of filaments and clumps. 
Two large spatial filaments were found to run in parallel over most of the MS
length.  
Around Dec $\sim0$\degree~ the dual filaments disperse into many small
clumps and filaments culminating in a chaotic appearance at the tip.
The only high resolution view of two selected regions at the tip of
the MS was by Stanimirovic et al. (2002) who used the Arecibo telescope 
to image two small regions in MS V (Dec $\sim8$\degree) 
and MS VI (Dec $\sim12$\degree). This work showed that the MS 
clumps have a complex morphology at 3.5$'$ resolution, strongly
suggestive of interactions between the MS and an external medium. 
While most previous studies thought that the MS dissipates 
at its tip (Dec $\sim0$\degree),
very sensitive Westerbork observations by \cite{Braun04b} suggested
that the MS remarkably extends further to the north
all the way to Dec $\sim40$\degree.

Another interesting phenomenon brought to light by the Parkes surveys
is the presence of numerous small HI 
clumps which surround the main MS filaments in position and velocity \citep{Putman03}.
While the origin of these clumps is still unclear, several possible mechanisms
have been invoked: the clumpiness in the original gas
drawn out of the Magellanic Clouds, the instabilities along the MS's edge, 
or dense condensations within an extended
mainly ionized MS component.
One of the crucial issues about the origin and structure of the MS in general,
is to what extent interactions with the MW halo determine or influence the
MS gas.  This problem becomes particularly important at the extreme northern
end of the MS, because 
this portion of the MS is considered to be the oldest and 
has been immersed in the hot MW
halo for a long time.

In this paper we present results from the recent HI observations of the tip
of the MS obtained with the Arecibo telescope\footnote{The Arecibo 
Observatory is part of the National
Astronomy and Ionosphere Center, operated by Cornell University under a
cooperative agreement with the National Science Foundation.} as part of 
the on-going survey by consortium for Galactic studies with the 
Arecibo L-band Feed Array (GALFA).
In Section 2 we briefly outline our observing and data processing
methods. Section 3 describes several new filaments discovered at the tip
of the MS, while in Section 4 we present a statistical summary of properties
of numerous HI clouds found in this region. 
In Section 5 we discuss the origin of the MS and its clumpy structure based
on  our observations and results.  

\section{Observations and Data Processing}
\label{s:obs}

The observations were conducted with  the Arecibo
telescope. GALFA HI survey consists of 
many individual projects. The data presented in this paper
represent a combination of three GALFA projects:
TOGS (or `Turn-On GALFA Spectrometer' being undertaken in
parallel with the ALFALFA extragalactic survey, PI: Putman), a2172 (PIs: Heiles \& Peek,
whose original target was a shell-like structure at Galactic velocities), and
a2032 (PI: Stanimirovic, which observed a small 
region at the MS tip).

As general GALFA observing and data reduction strategies are summarized in
\cite{Stanimirovic06}, we emphasize here only a few important points. 
All observations were obtained with
the dedicated spectrometer GALSPECT which has a fixed
velocity resolution of 0.18 \kms. 
For a2032 and a2172, the telescope was driven in the basket-weave mode
producing inter-woven zig-zag scans in the RA--Dec coordinate frame.
For TOGS, numerous drift scans along
right ascension were made, as this is the preferred observing strategy for 
the ALFALFA survey \citep{Giovanelli05}.
Before each observing scan
about 5 minutes were spent running  ``The Least-Squares Frequency Switching'' 
calibration procedure \citep{Heiles07} which allows us to derive a reference spectrum
to be applied to all spectra within the given scan. 

We have then combined scans from TOGS (the `Fall' session) and a2172 to obtain
the data cube that covers a region $37.5$\degree $\times24$\degree~in size. 
Various zig-zag and drift scans cross in many places. The crossing points
were used to compare spectra and fine-tune our calibration.   
This procedure removes 
small gain offsets between adjucent drift scans and produces 
a resultant image mostly free of stripping artifacts.
After this fine-tuning, 
spectra from the two projects were gridded into a 
 data cube.
In the case of a2032 observations our strategy was different.
After basic calibration, we have produced an a2032 data cube,
and then added it linearly (in the image domain) to 
the combined TOGS$+$a2172 cube.
Both combination processes worked very well.
The images presented in this paper are the first demonstration of
how successfully we can  piece together data from different GALFA projects.

The only problem encountered during the data reduction
is radio interference present is several TOGS scans at high negative LSR velocities.
This interference affects all ALFA beams, varies
across a given scan, and is centered around 1422 MHz.
It is caused by a still unidentified 
source. We have edited affected scans where possible, however
some artifacts are still noticeable in several images.
Our future software developments will be better able to deal with this, and similar
problems.

As the integration time was different for the three observing projects, our
sensitivity varies across the image. ALFALFA and TOGS cover the sky twice 
with about 15 sec of integration time per beam, while
a2172 and a2032 observations were obtained in a fast-scanning mode
with an integration time of only about 5 sec per beam.
Declination bands from 11\degree~to 16\degree~and 24\degree~to 28\degree~have
the lowest rms noise level, 0.03 K per 1.4 \kms~wide channels.
The rest of the image is noisier, with about 0.06-0.07 K per 1.4 \kms~wide channels. 
Our best 3-$\sigma$ sensitivity is $3.3\times10^{18}$ cm$^{-2}$ over 20 \kms.

\section{Four filaments at the tip of the MS}
\label{s:results}

\begin{figure*}
\epsscale{1.3}
\plotone{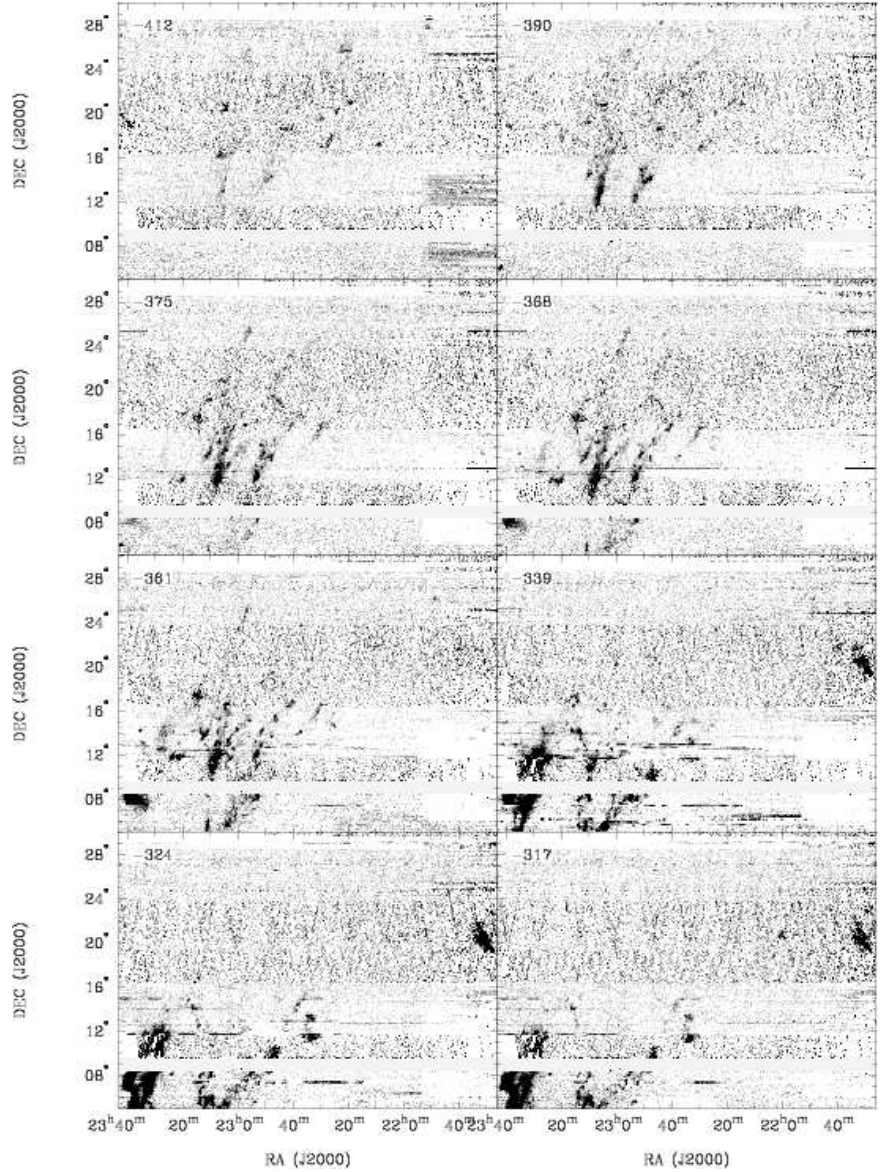}
\caption{\label{f:chans} Right ascension-declination images of HI emission at
  the tip of the MS for differentLSR velocities, given in the top left corner of
  each panel.  The gray-scale intensity range is 0.005 to 0.15 K, with a
  linear transfer function. Several horizontal strips around Dec $=6$\degree
  and $13$\degree, especially noticeable at lower negative velocities, are due
  to interference. The gap between Dec $\sim8$\degree~and 10\degree~is due to
lack of data in this region. The zig-zag pattern noticable in regions with a higher
noise level is due to our scanning technique.}
\end{figure*}

\begin{figure*}
\epsscale{1.2}
\plotone{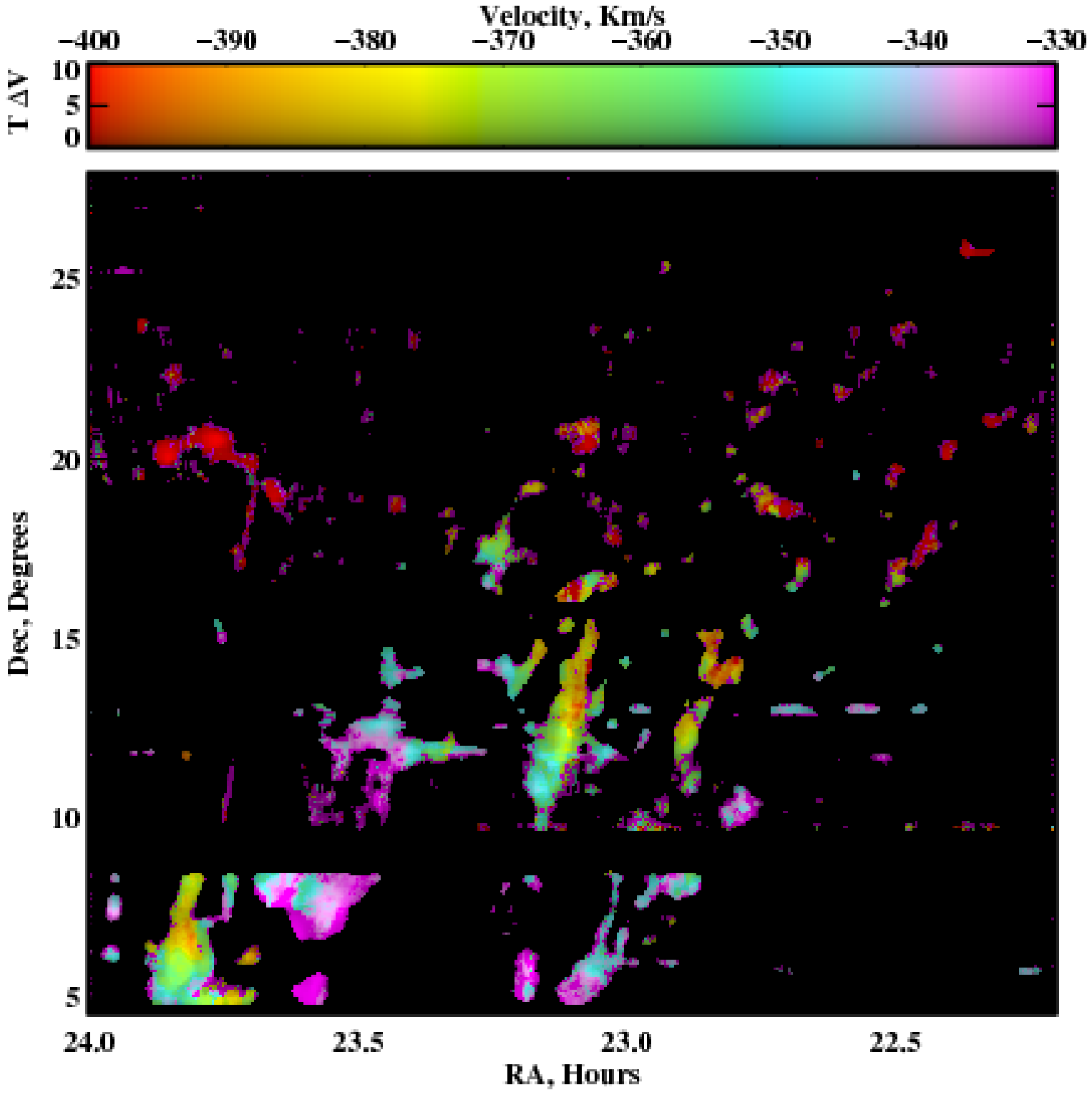}
\caption{\label{f:moments} First moment image, or the velocity field, of the MS
  tip.  
   Color represents the velocity centroids and brightness represents
integrated intensity. The image has been smoothed to emphasize weak features.}
\end{figure*}

Figure~\ref{f:chans} shows HI images from the final data cube at several selected velocities.
In Figure~\ref{f:moments} we show the first moment image, or the intensity-weighted
velocity along the line-of-sight. 
The area covered in our observations is huge --- about 870 square degrees  
($65$\degree$<l<$ 105\degree, $-54$\degree$<b<-20$\degree), focusing on the
very tip of the MS where the two traditional MS filaments were found to be significantly 
spatially bifurcated (one is located at RA 23$^{\rm h}$ 30$^{\rm m}$, 
and the other one at RA $\sim$23$^{\rm h}$ 00$^{\rm m}$). 
The only previous high resolution observations of about 10 square degrees
within this region were obtained by Stanimirovic et al. (2002).

As Figures~\ref{f:chans} and \ref{f:moments} show, several coherent 
HI filaments extend all the way to 
Dec $\sim25$\degree. \cite{Braun04b} reported an extension of the two  MS 
filaments, in the form of diffuse gas, from Dec = 20\degree~to 40\degree. 
We confirm the continuous extension of the MS from Dec = 5\degree~to $\sim$25\degree.
Beyond this point our observations may not be sensitive enough to detect
a very low-column-density emission found by \cite{Braun04b}.
Our sensitivity of $3\times10^{18}$ cm$^{-2}$ (over 20 \kms)
is significantly worse than \cite{Braun04b}'s lowest 
contour (in their Figure 5) shown at $3\times10^{17}$ cm$^{-2}$. 
However, the picture of the two interwoven filaments changes dramatically 
in our observations: instead of two traditional filaments we find several!

\begin{figure*}
\epsscale{1.3}
\plotone{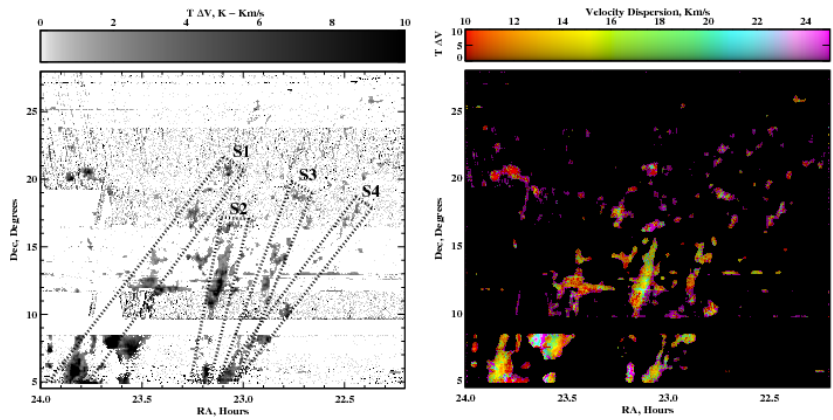}
\caption{\label{f:intensity_ann} (left) An HI integrated intensity image with schematic
  representation of the four filaments. (right)
Second moment or velocity dispersion.}
\end{figure*}

While filaments often break into numerous branches and 
small clouds when viewed with Arecibo's resolution,
there are at least four prominent and coherent large-scale structures 
noticeable across the whole velocity range. To facilitate further discussion
we define and label filaments in the following way.
Starting from the lowest negative velocities ($\sim -320$ \kms) in
Figure~\ref{f:chans}: 
\begin{enumerate}
\item `S1' is centered at RA 23$^{\rm h}$ 35$^{\rm m}$, this is a portion of one of
the two traditional MS filaments;  
\item `S2' is centered at RA 23$^{\rm h}$ 10$^{\rm m}$ and can be traced through the
whole velocity range;
\item `S3' is centered at RA 23$^{\rm h}$ 05$^{\rm m}$ and starts at a higher negative 
velocity of $-360$ \kms;
\item `S4' starts at RA 23$^{\rm h}$ 00$^{\rm m}$ and extends to RA 22$^{\rm h}$ 35$^{\rm m}$.
\end{enumerate}

Figure~\ref{f:intensity_ann} (left) shows a schematic representation of the
four streams.
Both S1 and S4 can be seen from the lowest negative velocities
up to about $-370$ \kms.
S2, S3 and S4 seem to have spatially a common point of origin, and 
are most likely related to the second traditional MS filament.
The morphology of the four streams is significantly different:
S2, S3 and S4 consist of and are surrounded by numerous small, compact
clumps, having a beads-on-string appearance.  
The number of compact clumps peaks at velocities $-350$ to $-370$ \kms.
S1 on the other hand has a more diffuse morphology and contains the smallest 
number of compact clumps.
This may indicate that S1 has a different origin relative to the other three streams, 
or that it is younger and therefore did not have 
enough time to fragment. We get back to this issue in Section~\ref{s:discussion}.

\begin{figure*}
\centering
\includegraphics[width=.7\textwidth]{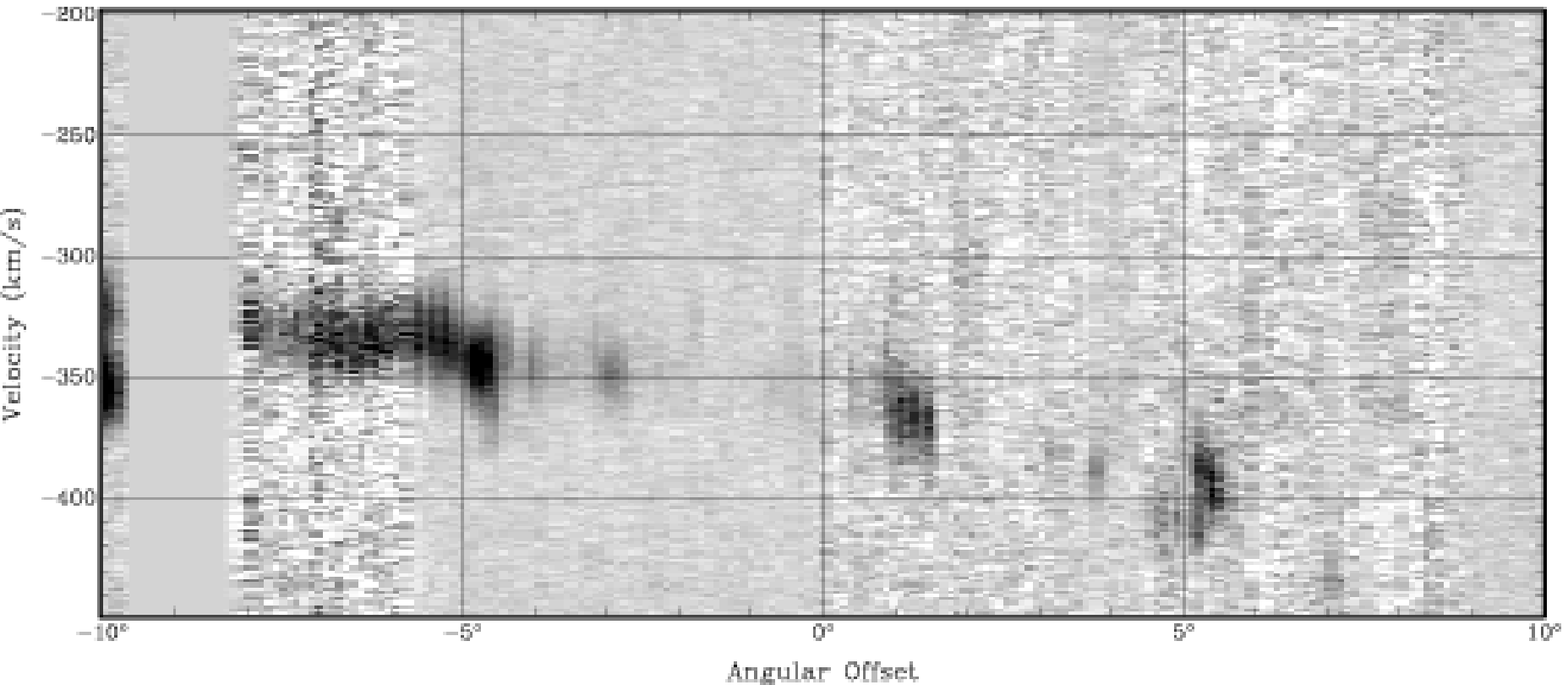}
\includegraphics[width=.7\textwidth]{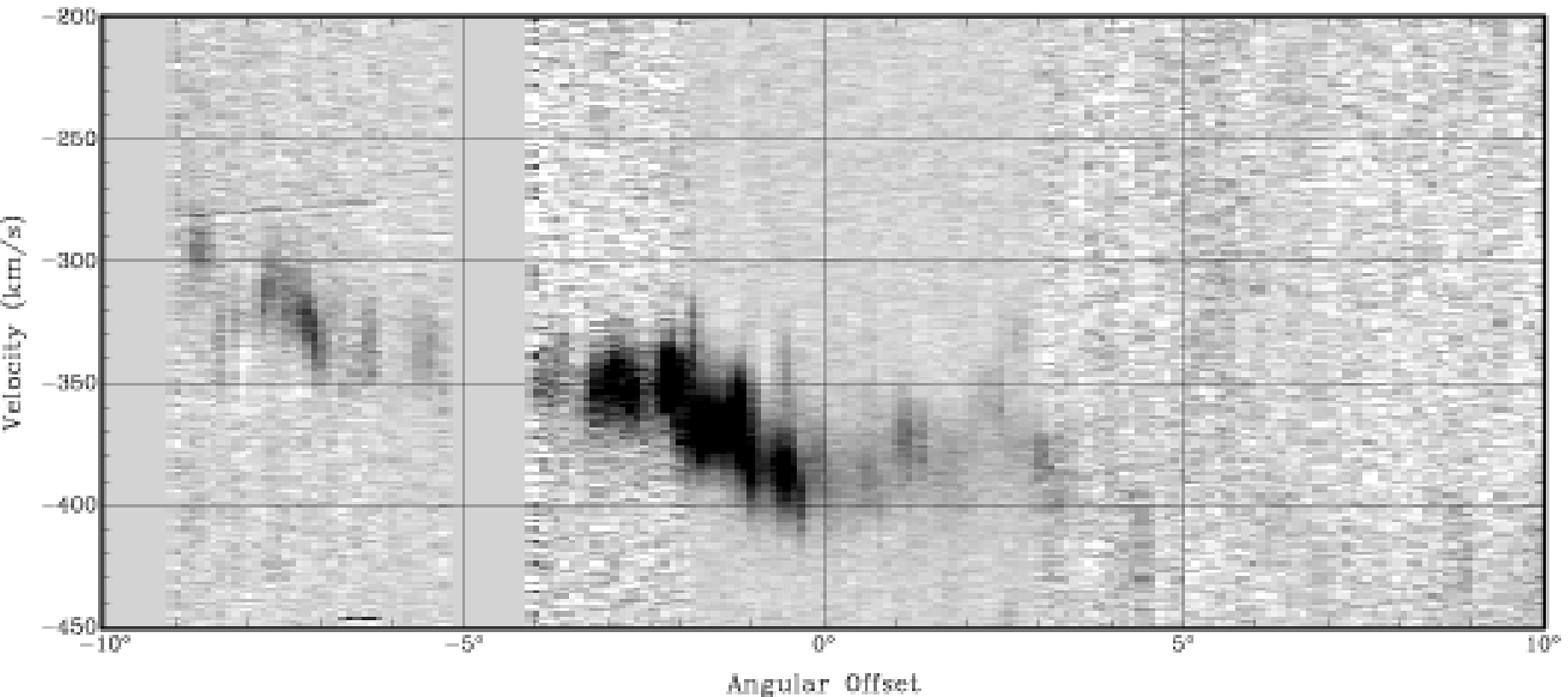}
\includegraphics[width=.7\textwidth]{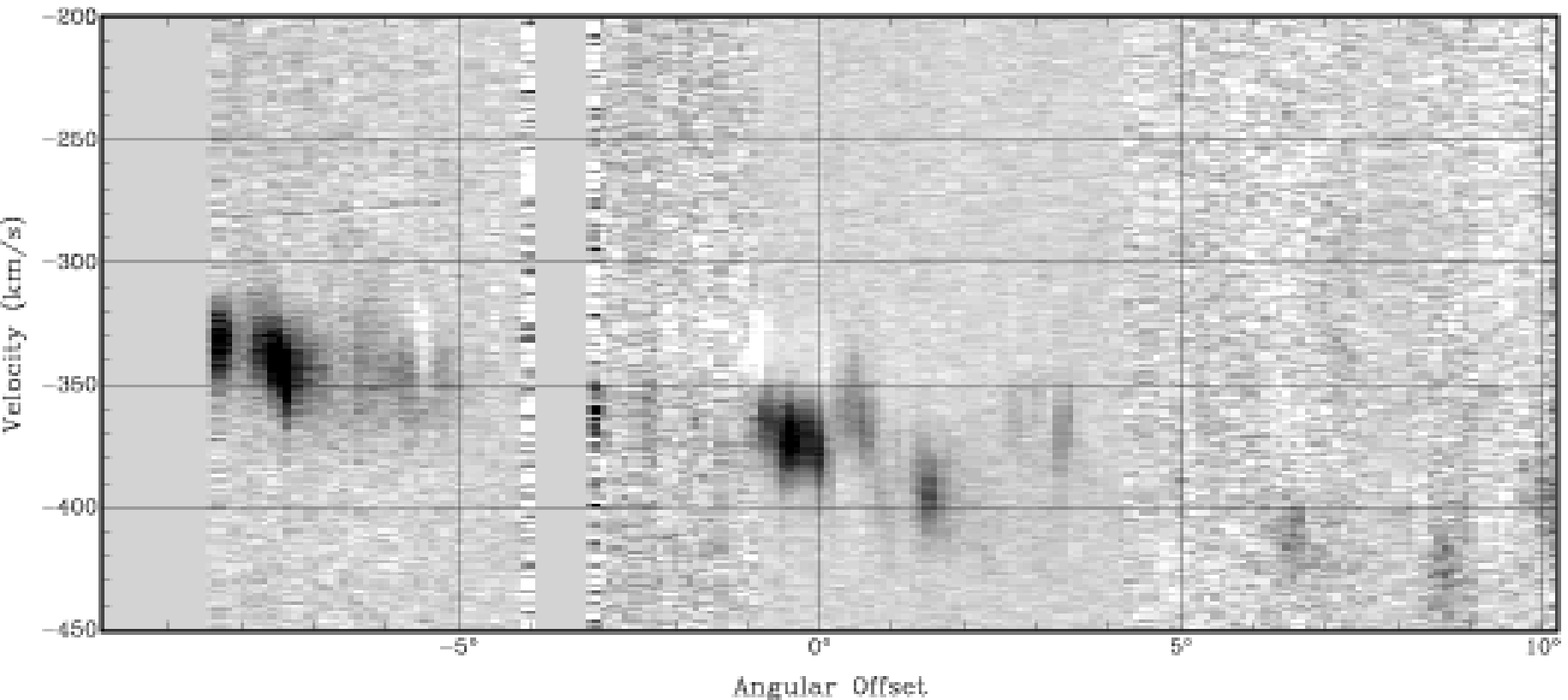}
\includegraphics[width=.7\textwidth]{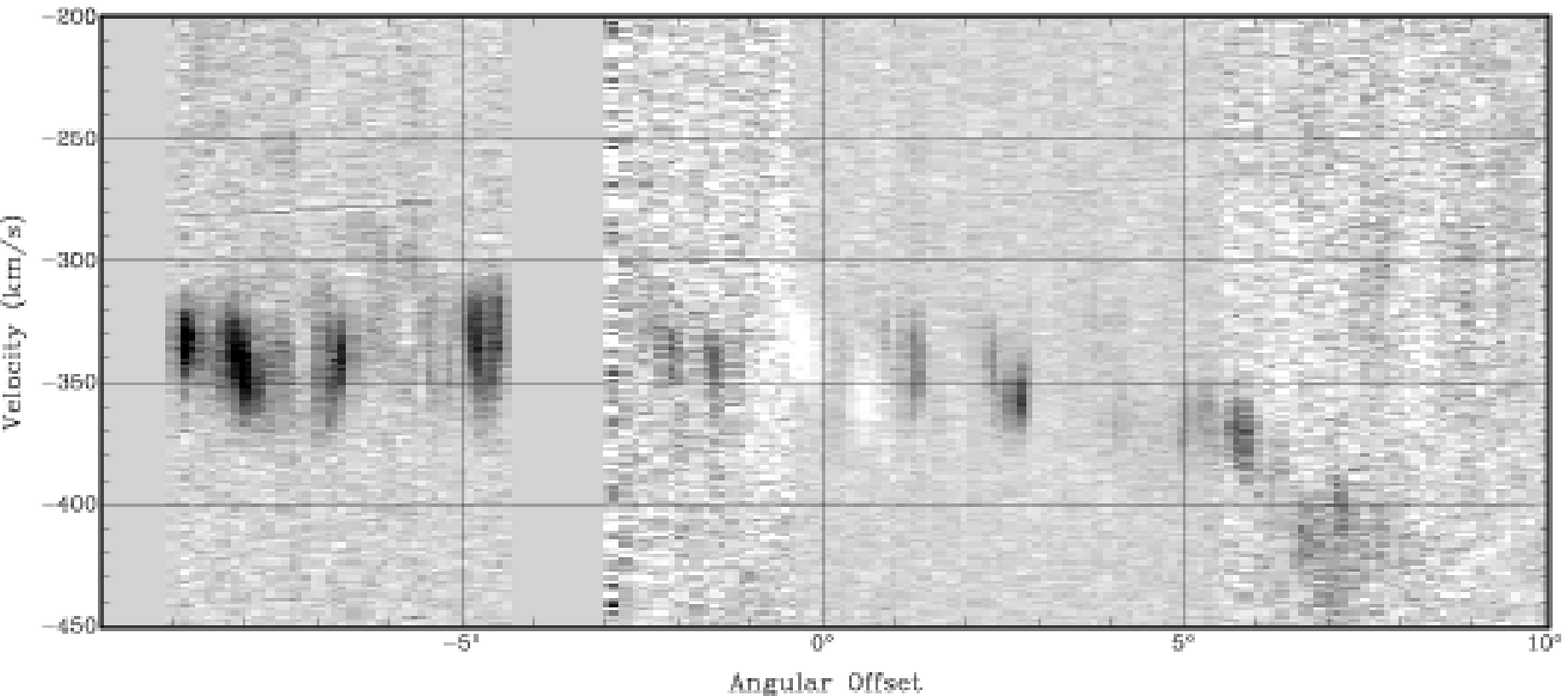}
\caption{\label{f:cuts} Position-velocity images along the four
  streams: S1, S2, S3, and S4 (top to bottom).
This figure can be compared with Figure 9 from Connors et al. (2006).
}
\end{figure*}

\begin{figure*}
\epsscale{1.3}
\plotone{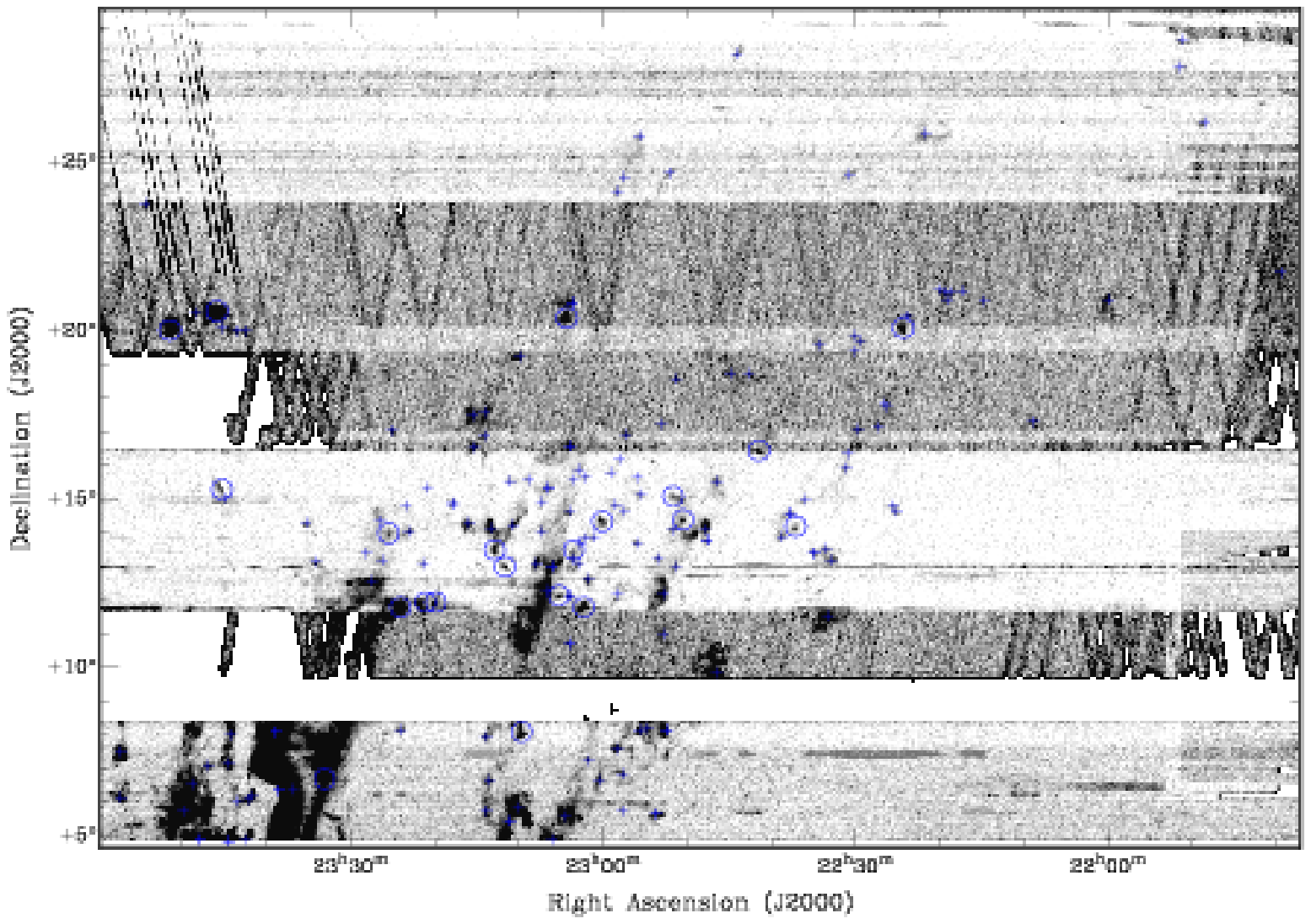}
\caption{\label{f:all_clouds}
A peak brightness temperature image of the whole region with cataloged clouds.
Clouds without multi-phase structure are shown as crosses, while clouds
with multi-phase structure are shown as circles.  See Section 4.2.}
\end{figure*}

While generally clumpy morphology is prominent, 
the four defined streams  are coherent structures,
which can be seen most clearly in the first moment image,  Figure~\ref{f:moments}.
All four streams show clear velocity gradients, although not all gradients are
similarly steep. 
In Figure~\ref{f:cuts} we show position--velocity images along the
streams.
S1 and S3 have a velocity gradient of about 5 \kms deg$^{-1}$ over a length of 15
degrees, in agreement with the overall velocity gradient along most of the
MS \citep{Putman03}.
S2 has the steepest gradient of almost 10 \kms deg$^{-1}$ 
along the first 10 degrees, and then it almost turns over in velocity 
roughly staying around $-380$ to $-400$ \kms. 
S4 has almost a constant velocity of $\sim-340$ \kms.
All streams have a similar velocity dispersion, $\sim15$-20 \kms,  
see  Figure~\ref{f:intensity_ann} (right).
This is in agreement with \cite{Bruns05} 
who measured a dispersion of 15 \kms~along most of the MS.

\section{Compact HI clouds at the tip of the MS}

\begin{figure*}
\epsscale{1.2}
\plotone{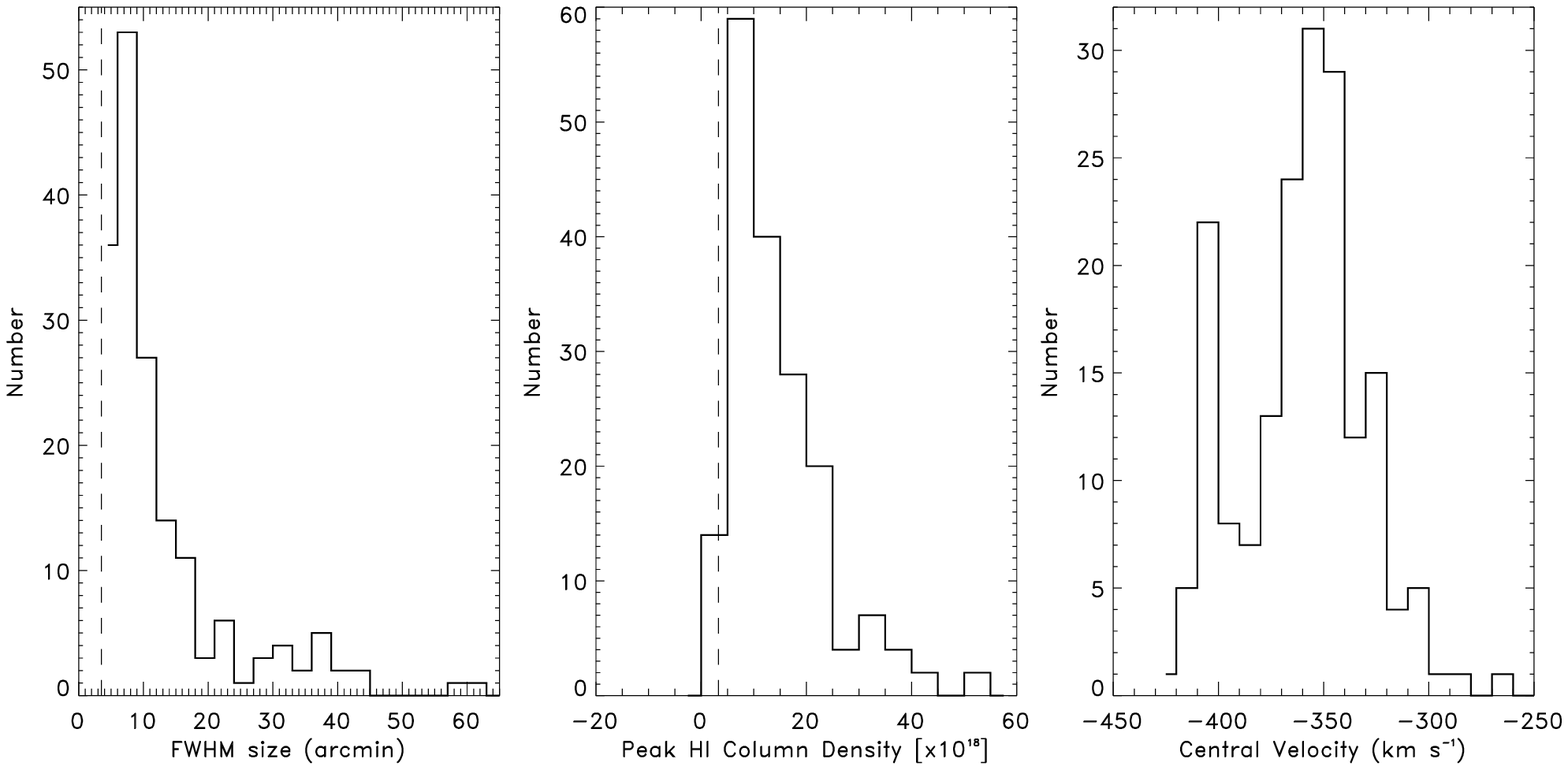}
\caption{\label{f:cloud_histos} Histograms of measured cloud properties: cloud angular size
(FWHM in arcmin), the peak HI column density (in 10$^{18}$ cm$^{-2}$), and the central
LSR velocity. The dashed line in the left panel shows the angular resolution
limit, while the dashed line in the middle panel shows the 3-$\sigma$ sensitivity limit
of the survey. }
\end{figure*}

\begin{figure*}
\epsscale{1.2}
\plotone{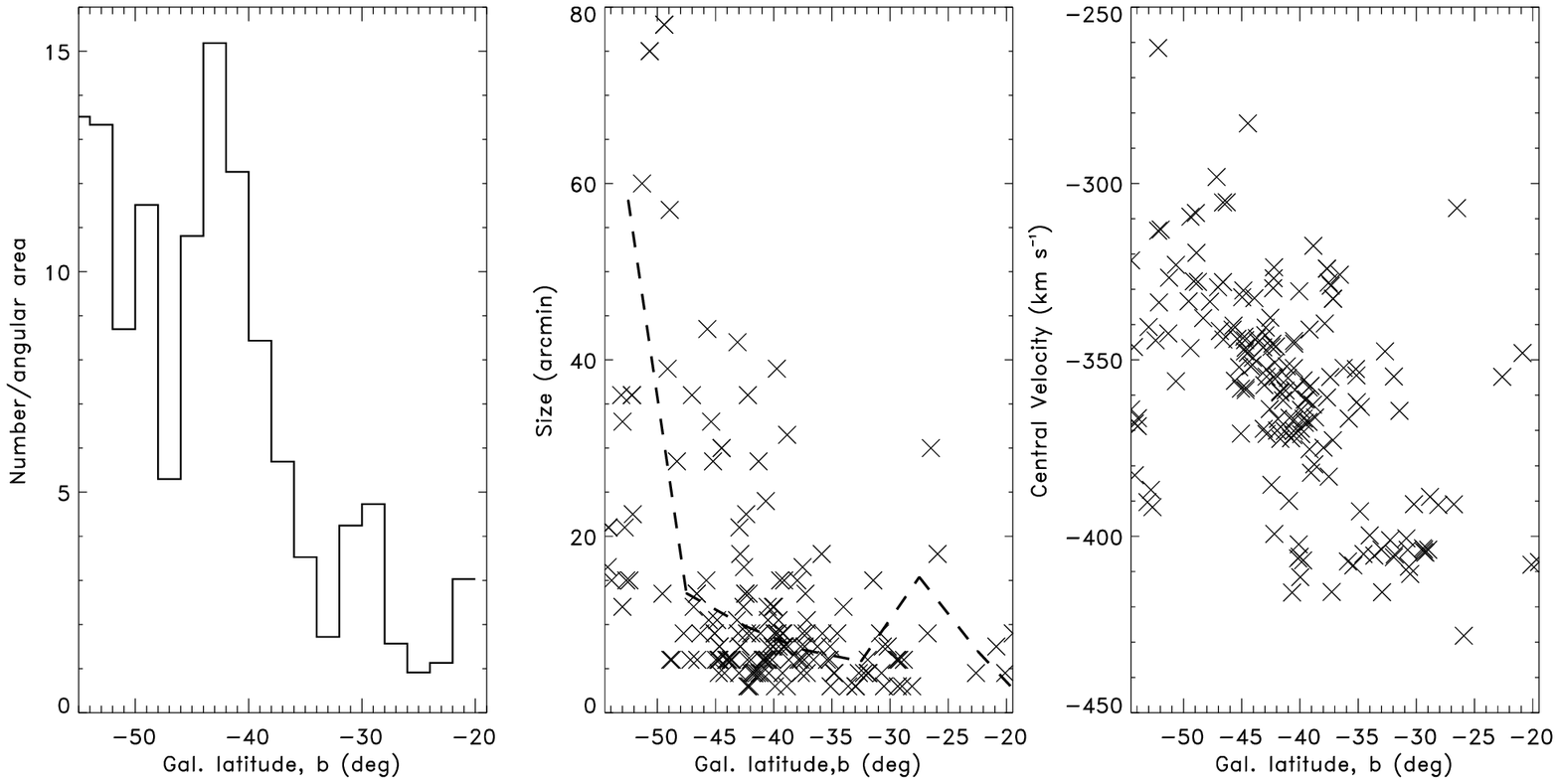}
\caption{\label{f:dec_vel_more}
(left) Number of clouds as a function of Galactic latitude
divided by the angular area observed at each latitude. 
(middle) Cloud size as a function of Galactic latitude. 
Dashed line represents a median value for each latitude bin.
(right) Cloud central velocity as a function of Galactic latitude.
 In the case of clouds with multi-phase structure,  the mean velocity
 is plotted.}
\end{figure*}

A striking result from the above figures is the presence of numerous small HI clouds.
The smallest clouds have a size down to our resolution limit of 3.5$'$ and are unresolved. 
Most of the clouds are compact and round, head-tail morphology is surprisingly uncommon.
Previously, \cite{Putman03} noticed two types of clouds which surround the MS along 
its southern portion: 
head-tail clouds with tails pointing along the major axis of
the MS (typically, the head has a column density of $10^{19}$ cm$^{-2}$ while
the tail is 3-4 times fainter),
and ubiquitous dense HI clouds which follow the main MS filaments in position
and velocity, some again with a head-tail morphology.

\subsection{Cloud catalog}

We have compiled a catalog of 180 HI clouds. 
The positions of all clouds are shown in Figure~\ref{f:all_clouds}.
Clouds were selected by eye from the spectral line data cube. 
The integrated intensity image was used as a guide for cloud selection,
local peaks in this image were considered as potential cloud candidates.
We then used the full data cube to follow potential clouds spatially
and in velocity, and measure cloud properties.
Special care was taken to avoid breaking larger clouds into separate
entities, however the catalog may still be slightly 
biased toward smaller clouds.  
We have made several checks to ensure cloud reliability and to 
exclude clouds potentially affected by scanning artifacts or RFI.
There were several cases (mainly at low Dec) where multiple 
clouds are present along the line-of-sight at clearly different velocities. 
These cases were treated as individual clouds. 

For each cloud
we have measured the cloud size, the central velocity, and the
peak HI column density. In addition, at the center of each cloud we have 
fitted the HI velocity profile with one or two Gaussian functions.
Although the
cloud catalog is not 100\% complete, as it is hard to select clouds 
in the regions with higher noise, it certainly represents well the cloud
population whose physical properties we want to study.
Basic cloud properties are presented and discussed in the following sub-sections.

To ensure that we are dealing only with clouds of Magellanic origin,
our cloud catalog was cross-correlated with the catalog of HVCs by \cite{Wakker91},
the catalog of HVCs and CHVCs by \cite{deHeij02}, and the catalog
of recently discovered mini-HVCs by \cite{Hoffman04}.
HVCs come with a wide range of sizes and properties. While some represent
large complexes, like the MS, most are smaller with sizes of a few tens of
degrees. On the other hand, CHVCs represent a class of HVCs which are compact 
(with sharp edges in the HI
column density distribution), apparently isolated 
from larger complexes, and
with angular size less that 2\degree.
One interpretation is that CHVCs are essentially the same as HVCs but located 
at much greater distances. 
Recently discovered mini-HVCs are even smaller and have diameters of 9$'$ to
35$'$. These clouds are also 
more diffuse than typical HVCs, having the peak HI column density 
of only  $3\times10^{18}$ cm$^{-2}$.
Several of mini-HVCs were found to be
superimposed on the northern extension (Dec $>0$\degree)
of the MS but at velocities distinct from those of the MS gas. One
suggestion is that these clouds represent outliers of the MS 
complex \citep{Hoffman04}.

There are 8 HVCs from \cite{Wakker91} in our field of view. 
Three HVCs are clearly detected and identified as clouds in our catalog, these are
WvW413, WvW491 and WvW485. Additional three HVCs 
(WvW508, WvW492 and WvW472) have clouds nearby but at an angular 
separation larger than the beam of the Leiden-Dwingeloo survey.
There are 19 HVCs and 3 CHVCs from \cite{deHeij02} in our field of view.
All of these CHVCs were detected. 

As CHVCs were classified as isolated clouds it will be surprising to
find a large population of CHVCs in our cloud catalog.
Only 3 CHVCs from
\cite{deHeij02} are in our field of view. 
CHVC229 (from de Heij et al. 2002) is extended and was found 
to be associated with several of our clouds at very
low negative velocities ($\sim-330$ \kms). We have excluded those
HI clouds from our further analyses as at $\sim-330$ \kms~these clouds 
are clearly spatially located far away from the MS emission.

Three mini-HVCs are in our field of view: LMPU 387-230, LMPU 387-275, 
and LMPU 387-297. LMPU 387-275 and LMPU 387-297 are at velocities 
($-275$ and $-297$ \kms, respectively) 
where we detect several small clumps, the closest one being about 24$'$ away.
We did not detect LMPU 387-230 however, although \cite{Hoffman04} estimated its
HI column density to be $4.8\times10^{18}$ cm$^{-2}$, clearly above our detection
limit.

In summary, we do not find that our catalog is contaminated by a significant
population of CHVCs, it therefore contains primarily clouds associated with the MS.

\subsection{Cloud properties}
\subsubsection{Cloud size, HI column density,  and HI mass}

\begin{table*}
\footnotesize
\caption{Cloud radius, HI mass, HI volume density, and thermal pressure.}
\centering
\label{table1}
\begin{tabular}{lcccc}
\noalign{\smallskip} \hline \hline \noalign{\smallskip}
Distance   & Radius    & HI mass & HI volume density& Pressure \\
(kpc) & (pc) &(M$_{\odot}$) & (cm$^{-3}$)& (K cm$^{-3}$)  \\
\hline
20 & 30 & $2\times10^2$& 0.4 & 40 \\
60 & 90 & $2\times10^3$& 0.1 & 10 \\
150 & 220 & $1\times10^4$ & 0.05 & 5\\
\noalign{\smallskip} \hline \noalign{\smallskip}
\end{tabular}
\end{table*}

Figure~\ref{f:cloud_histos} shows histograms of 
cloud basic properties: angular size, HI column density, and central (LSR) velocity.
The observed cloud angular size distribution strongly peaks around 10$'$. 
90\% of clouds have a size in the range 3$'$ to 35$'$.
The abundance of very small, compact-looking clouds is significant,
more than half of clouds in the catalog have an angular size less than
10$'$. Obviously, these clouds would have been missed in previous
large-scale surveys such as the Leiden-Argentine-Bonn (LAB) survey 
\citep{Kalberla05a,Hartmann97,Bajaja05,Arnal00}, or even Parkes surveys of the MS.
The low-size cut-off of the histogram is clearly affected by our
resolution limit of 3.5$'$, and higher resolution observations 
will probably find many smaller clouds.

The cloud central HI column density distribution peaks around $1.3\times10^{19}$
cm$^{-2}$, 90\% range is $2.5\times10^{18}$ to $3.2\times10^{19}$ cm$^{-2}$.
This distribution function is affected by our sensitivity limit, deeper
observations will probably find many clouds with lower HI column density.
For comparison, the HI column density distribution of mini-HVCs
peaks at $3\times10^{18}$ cm$^{-2}$.

Figure~\ref{f:dec_vel_more} (left) shows that the number of clouds decreases
steeply along the MS, in the direction toward the tip. The number of
clouds in each latitude bin has been divided by the observed angular area 
to account for different areal coverage across the image. 
Figure~\ref{f:dec_vel_more} (middle) shows a tendency for 
clouds with angular size larger than 20$'$ to be found closer 
to the Magellanic Clouds (at $b<-40$\degree), while clouds with angular size less
than 20$'$ appear along the whole latitude range.
This could be interpreted as evidence that 
part of the MS with $b<-40$\degree~is at a larger distance, 
in agreement with tidal models.
Please note that the gap around $b=-25$\degree~on this plot is due to the lack
of clouds, not due to the lack of data in this region.

To provide an example of cloud linear size, HI mass, volume density,
and thermal pressure, we use an angular diameter of 10$'$ and a 
median peak HI column density of $1.3\times10^{19}$ cm$^{-2}$, 
for three representative distance values, 20, 60, and 150 kpc, see 
Table~\ref{table1}.

\subsubsection{Cloud central velocity}

Interestingly, the cloud central velocity, shown in
Figure~\ref{f:cloud_histos} (right), 
shows a clear dichotomy. There are two peaks centered at $-406$ and $-351$
\kms, with the velocity FWHM of 10 and 45 \kms, respectively.
As shown in Figure~\ref{f:dec_vel_more} (right),
the velocity dichotomy manifests in the way that there are two main groups
of clouds, one around $b=-47$\degree~to $-37$\degree~(or Dec = 10\degree-15\degree) 
and $v=-350$ \kms, and the other one around $b=-35$\degree~to 
$-27$\degree~(Dec = 18\degree-25\degree) and $v=-400$ \kms.
This figure also shows an almost continuous linear decrease in
velocity (from $-300$ to $-420$ \kms) with Galactic latitude. 
This is in agreement with the general velocity trend along the MS.

The observed velocity dichotomy may 
indicate a kinematic bifurcation of the MS tip.
Such bifurcation would not be surprising. N-body simulations by \cite{Connors06} found
two kinematically separate components, one that follows closely the orbit
of the LMC and has a lower negative velocity, and the other 
one with a more negative velocity. 
While separated kinematically, both components occupy the same $(l, b)$ range.

\subsubsection{Clouds with multi-phase structure}

\begin{figure*}
\epsscale{1.5}
\vspace{-1cm}
\plotone{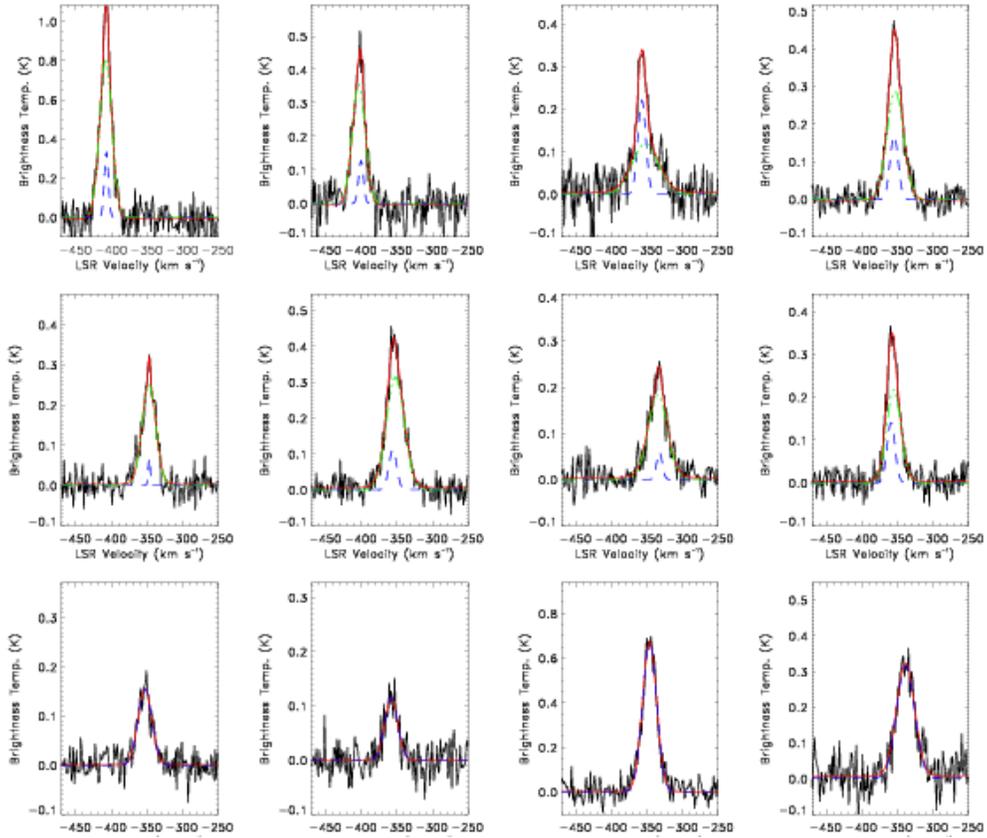}
\vspace{0.5cm}
\caption{\label{f:some_clouds}
A few examples of HI clouds with and without multiple velocity components. The
top two rows show clouds whose velocity profiles require two Gaussian
functions, the bottom row shows a few clouds with just a single Gaussian
velocity profile. For multiple velocity components, the broad 
velocity component is shown in green, the narrow 
component is shown in blue, and their sum is shown in red.}
\end{figure*}

About 20\% of the clouds in our catalog have  
velocity profiles whose fitting requires two Gaussian functions. 
This is indicative of the existence of a multi-phase medium.
However, some of these velocity profiles could be due to the existence of 
multiple clouds along the line of sight, whose velocity profiles partially overlap.
To account for this effect, we accept as multi-phase clouds only
those which have a subsonicly moving core and a warmer envelope.
This criterion translates into the requirement that 
the absolute difference between the centers of the two 
fitted Gaussian functions does not exceed $0.7\times$ the FWHM of the broad component. 
Our final sample of multi-phase clouds contains 21 entries.
As an example, in Figure~\ref{f:some_clouds} we show velocity profiles of ten
multi-phase (the first two rows), and five clouds with a single Gaussian profile
(the last row). In most multi-phase cases, a narrow Gaussian component 
fits the line center, while a broader Gaussian function is needed to 
fit the line wings. 

Narrow line profiles were scarcely found in the MS in previous studies.
\cite{Bruns05} found one cloud with a velocity FWHM of only 4 \kms, 
at $l=93.6$\degree, $b=-53.7$\degree.
\cite{Kalberla06} used the LAB survey and decomposed 
all HI velocity profiles into Gaussian components. They found that  
the portion of the MS close to the Magellanic Clouds (and with
positive velocities) has 27\% of the sight lines with a narrow velocity component,
while the portion further away from the Clouds (with negative velocities) 
has a significantly lower fraction, only about 10\%. 
The lower number of narrow components was attributed to either a 
lower halo pressure and/or a larger distance of the northern MS extension. 
Our final fraction of multi-phase clouds (12\%) is similar to 
Kalberla \& Haud's estimate for the negative velocity portion of the MS. 

\begin{figure*}
\epsscale{1.3}
\plotone{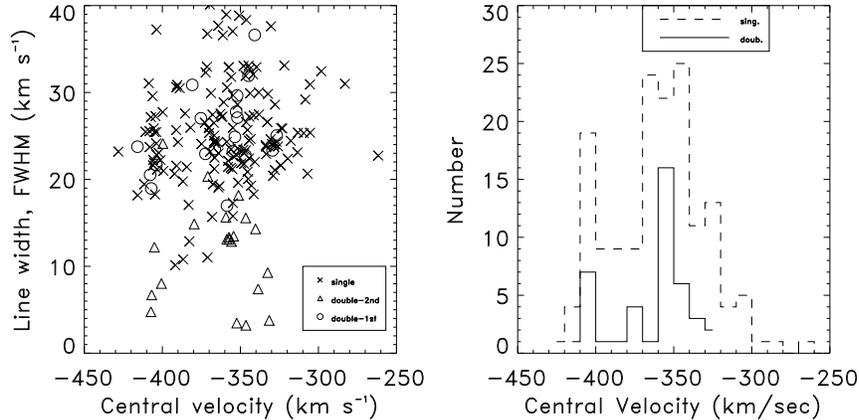}
\vspace{0.5cm}
\caption{\label{f:cloud_histos1} (left) Velocity FWHM vs central velocity for all
fitted Gaussian functions. Crosses correspond to clouds with only one
Gaussian component. 
Diamonds and circles show narrow and broad velocity component, respectively,
in the case of clouds with multi-phase structure. 
(right) Histogram of central velocities for clouds with a single Gaussian
function (dashed line), and clouds with 
two Gaussian functions (solid line). Central velocities of both Gaussian components
were used to derive the solid-line histogram.   }   
\end{figure*}

Figure~\ref{f:cloud_histos1} (left) shows the FWHM vs central velocity of all fitted
Gaussian functions: crosses correspond to clouds with a single Gaussian 
component, while diamonds and circles represent the secondary (or narrow) and
the primary (or broad) velocity component for clouds with multi-phase medium.
The FWHM of single-component clouds is predominantly in the range of 20--30 \kms.
For multi-phase clouds, the FWHM of the narrow component 
has a broad, almost continuous, range from 3 to 20 \kms, with 
a median value of 13 \kms.
This would imply an upper limit on the kinetic temperature
in the range 200--9000 K, mostly being in the thermally unstable regime.
For comparison, \cite{Kalberla06} find the FWHM of the narrow 
component in the range 5--28 \kms.
The FWHM of the broad component of multi-phase clouds is around 25 \kms, in
agreement with what is found for clouds with only a single Gaussian component.



The right panel in Figure~\ref{f:cloud_histos1} shows the
central velocities for both types of clouds.
The velocity dichotomy is still noticeable with 
two peaks around $-405$ and $-350$ \kms.
This shows that the velocity bifurcation affects both 
clouds with and without multi-phase structure. 
We return to this point in Section~\ref{s:distances}.

\section{Discussion}
\label{s:discussion}

\subsection{Origin of the MS filaments}

There are several important observational points presented in the previous
sections:
\begin{itemize}
\item there are four coherent streams at the tip of the MS;
\item S1 is diffuse, while S2, S3 and S4 are clumpy;
\item spatially, S2, S3 and S4 appear to originate from a similar location;
\item the velocity gradient along streams gradually decreases from S2 to S4; 
\item there is a lack of large clouds at low Galactic latitudes;
\item central velocities of clouds with and without multi-phase structure 
show velocity bifurcation at $-350$ and $-405$ \kms.
\end{itemize}

Traditionally, it was assumed that the MS originated purely from the SMC gas.
\cite{Putman03} suggested that the 
two double-helix-like MS filaments could arise from gas stripped from the
Magellanic Bridge and the SMC.
By using the LAB survey of Galactic HI
(angular resolution of 0.6\degree),
\cite{Nidever07} were able to trace
one of the two filaments all the way to an over-density region in the LMC.

The only numerical study that addressed spatial 
bifurcation of the MS is by \cite{Connors06}.
These authors 
showed that gross features of the MS and the leading arm, as
well as the spatial bifurcation of the main MS filament, 
can be reproduced purely by tidal interactions between 
the MW and the Magellanic Clouds.
They also suggested that the main MS filament
is broken into two kinematic components.
Several distinct features are found in their simulation.
\begin{enumerate}
\item The main MS filament was created 1.5 Gyr ago in a major encounter
between the SMC, the LMC, and 
the MW. Further tidal `kicks' from encounters with the  LMC 1.05 and 0.55 Gyrs ago
resulted in its spatial, and then kinematic bifurcation.
While one of the bifurcated components follows the orbit of the Magellanic Clouds and
has a lower negative velocity ($\sim-350$ \kms), 
the other possesses a higher negative velocity
($-500$ to $-400$ \kms).
\item 
In addition to the main stream, which extends in distance from 60 to $\sim100$ kpc, there is
the third stream right at the tip which spatially coincides with the main stream but
has distinct kinematics and is significantly further away (distance of 170 to
220 kpc).
This component was created in an encounter between the SMC and 
the LMC 2.2 Gyrs ago, and is
the oldest and the most distant feature in the Magellanic System.
\item Two tidal tails were drawn $<200$ Myr ago from the Magellanic Bridge and 
follow the main filament along most of its length.
\end{enumerate}

By drawing a parallel with this simulation, 
streams S2, S3 and S4 could represent a 3-way splitting of the main MS filament. 
As we show in the next section,
this gas has had enough time to cool and 
fragment and is therefore very clumpy. 
As suggested by the simulation, most of the spatial 3-way splitting could 
have occurred in the more recent past, 1 Gyr to 500 Myr ago, during close 
encounters between the MS and the LMC. 
The fact that S2 to S4 show a decreasing steepness in their velocity gradients
may be suggesting that these filaments were separated in subsequent
encounters.
S1 on the other hand, may represent one of the more recently formed tidal
tails, drawn out of the Magellanic Bridge $<200$ Myrs
ago. The fact that it is significantly less clumpy than the other three streams
suggests that S1 may be significantly younger, and have not had enough time 
to cool and fragment.

The lack of large HI clumps at the higher Galactic latitude
suggests that the far tip of the MS may be more distant.
However, the fact that we find clumps with multi-phase
structure almost along the whole latitude range implies that
the far tip is probably not too distant. As discussed in the next section,
it would be hard to justify the existence of multi-phase medium
at a distance of $\sim200$ kpc.
Another disagreement with the simulation concerns the observed
LSR velocity at the far tip, we find $\ga -420$ \kms, while the
simulation predicts velocities down to $-500$ \kms.

However, the
observed large-scale filaments at the tip of the MS,
as well as the evidence for kinematic bifurcation, give strong
support to the tidal formation scenario for the MS. 
While the observations are only qualitatively in agreement with the simulations,
and many details still have to be worked out, 
it is extremely encouraging to see that detailed spatial and velocity structure 
of the MS can be reproduced {\it purely} by tidal interactions.
Better observed coverage of this region, as well as high resolution
observations of the southern MS extension, will certainly help to
constrain this and other possibilities in the future.

\subsection{Origin of Compact clouds}

At certain velocities the MS has a very clumpy morphology. 
One of the obvious questions is: what physical processes are responsible for
such a clumpy
structure? 
In trying to understand the cloud origin  
we look briefly into three important physical processes that 
would affect a tidal stream of warm gas moving through a hot ambient medium. 
All three processes have their characteristic time and spatial scales.

\subsubsection{Thermal instability}

Thermal instability arises due to gas cooling and 
results in spatial fragments of the WNM,  
whose size is defined by the cooling length $\lambda_{\rm cool}$.
This process operates on the cooling timescale,
defined by $\tau_{\rm cool}=kT_{\rm w}/(\Lambda n_{\rm w})$, 
with $\Lambda$ being the cooling
function, $k$ being the Boltzmann constant, and $T_{\rm w}$ and $n_{\rm w}$ being the
temperature and volume density of the WNM gas. 
If we consider a WNM gas that originated in the SMC, then
$n_{\rm w}=5\times10^{-2}$ cm$^{-3}$ (based on the size and total mass of the SMC, also
measurements in \citep{Stanimirovic99}),
then for $T_{\rm w}=8000$ K we estimate $\lambda_{\rm cool}\sim100-200$ pc and
the cooling timescale of $\tau_{\rm cool}\sim20-30$ Myr.
For a lower value of $n_{\rm w}$, which is not unreasonable
for the gas drawn out of the SMC outskirts,
 both $\lambda_{\rm cool}$ and $\tau_{\rm cool}$
will increase.
Considering that most models suggest the MS age of 1-1.5 Gyr,
thermal instability, with a typical timescale of a few tens 
to a few hundreds of Myrs,
most likely have had a significant effect on the evolution
of the MS gas.

Recent numerical simulations by \cite{Audit05} 
provide an example of thermal condensation of warm gas.
These simulations focus on a collision of incoming turbulent WNM flows and
thermal condensation of WNM into colder gas.
While not exactly applicable in our case, it is still very instructive 
to notice the properties of typical thermal fragments.
A collision of incoming WNM streams creates
a thermally unstable region, which partially condenses into cold clouds.
The thermally unstable warm gas (with temperature of 200 -- 5000 K) 
has a filamentary morphology and 
its fragmentation into cold clouds (with temperature $<200$ K) 
is heavily controlled by turbulence.
Cold clouds formed in simulations are thermally stable and long-lived, 
in the case of stronger turbulence they have round morphology, 
while a weaker turbulence is responsible for more elongated morphology of cold clouds.
The simulations show that the fraction of cold gas (with temperature $<200$ K)
produced in this process ranges from 10\% in a strong turbulent case, 
to about 30\% in a weak turbulent case. 
It is interesting to note that 
the lower bound on this CNM-to-WNM fraction agrees with
the fraction of HI clouds with
cool cores we find in our observations (12\%).

\subsubsection{Kelvin-Helmholtz (KH) instability}

Kelvin-Helmholtz (KH) instability occurs at the interface of
the moving warm stream of gas and the ambient medium and also results in the
formation of small fragments with a power-law size distribution.
The timescale on which this instability develops, $\tau_{\rm KH}$ 
(given in Gyr), is given by: 
\begin{equation}
\tau_{\rm KH} =\frac{R_{\rm cl} D_{\rm n}^{1/2}}{v}
\end{equation}
\citep{Murray93}, where $R_{\rm cl}$ would be the initial size (radius) of the 
MS gas pulled out of the Magellanic Clouds, 
$D_{\rm n}=n_{\rm w}/n_{\rm h}$ corresponds to the density contrast of the warm KH fragments
and the ambient hot medium, and $v$ is the stream velocity.
We use $v=220$ \kms~\citep{Stanimirovic02}, assuming that the velocity of
the original MS gas is equal to its present day velocity.
We also assume that $R_{\rm cl}\sim10$\degree, based on the size
of the MSVI portion of the MS, and that this material
was initially at a distance of 60 kpc.

If $n_{\rm h}=10^{-4}$ cm$^{-3}$, and $n_{\rm w}=5\times 10^{-2}$ cm$^{-3}$,
then $\tau_{\rm KH}\sim600$ Myr. 
As this timescale is almost comparable to the age of the MS (in the
framework of tidal and ram-pressure models),
we can conclude that the KH instability has had 
a negligible effect on the evolution of the MS gas.

\subsubsection{Cloud evaporation}

Once the warm stream has fragmented due to thermal/KH instability into smaller clouds,
the third process becomes important --- this is conductive heat transfer
between the warm stream fragments and the hot ambient medium.
\cite{Lin00} show that the size of the smallest fragments is set
by the cloud ability to radiate away the energy flux due to
conduction.

The warm clouds reside within warmer phases, hot halo gas 
with $T_{\rm h}\sim10^{6}$ K and warm ionized
gas with $T_{\rm h}\sim10^{5}$ K. 
Detections of OVI absorption from gas associated with the MS 
\cite{Sembach03} give strong support for the existence of an ionized 
component around the MS with $T_{\rm h}<10^{6}$ K.
It is expected that different phases are
separated by an interface region through which heat flows between the
two phases. The interface is a surface of transition where, for small
clouds, the WNM evaporates into the warmer medium.  
The classical evaporation theory
\citep{McKee77b}  predicts a critical radius for clouds at which
radiative losses balance the conductive heat input, and at which the
cloud neither evaporates nor condenses. Clouds smaller than the critical
radius evaporate, while clouds larger than the critical radius accrete
by condensation of the surrounding gas.

In the case of an interface between the WNM and the HIM,
the cloud
critical radius ($R_{\rm rad}$) times the density of the surrounding medium, $n_{\rm h}$, 
is $5\times10^{17}$ cm$^{-2}$ for $T_{\rm h}=10^{6}$ K, or
$5\times10^{16}$ cm$^{-2}$ for $T_{\rm h}=10^{5}$ K
(from Fig. 2 in McKee \& Cowie 1977).
Generally, the expected critical 
radius decreases with increasing $n_{\rm h}$.
However, for $n_{\rm h}\sim10^{-4}$ cm$^{-3}$, 
the critical radius is large, $R_{\rm rad}\ga200$ pc. 
Clouds smaller  than $\sim200$ pc would then evaporate, while WNM clouds larger 
than $\sim200$ pc would accrete the HIM by condensation. 
Thermal fragments we considered in Section 5.3 with a size 
$\lambda_{\rm cool} \sim200$ pc will be predominantly smaller than the critical 
radius and therefore undergoing evaporation.

These small ($\sim200$ pc), evaporating clouds are surprisingly long-lived though.
For mainly neutral clouds surrounded by the hot gas,
the classical mass-loss rate is $2\times10^{29}$ gr yr$^{-1}$ for $T_{\rm h}=10^{6}$ K, or
$\sim10^{27}$ gr yr$^{-1}$ for $T_{\rm h}=10^{5}$ K \citep{Slavin07}.
In the case of saturated evaporation the mass-loss rates are even lower.
For $n_{\rm w}=5\times10^{-2}$ cm$^{-3}$ 
the evaporation timescale is $\tau_{\rm evp}=0.4$ Gyr for $T_{\rm h}=10^{6}$ K, or
$\tau_{\rm evp}=80$ Gyr for $T_{\rm h}=10^{5}$ K.
But the timescales are in general long, approaching or being
longer than 1 Gyr.
We hence conclude that small fragments
would be able to survive for a long time, at least a Gyr, most likely throughout the whole
lifetime of the MS.

\subsubsection{Constraining cloud distances}
\label{s:distances}

As thermal instability must have played a very important 
role in the evolution of the oldest MS gas, we suggest that this could be 
the dominant process responsible for the clumpy structure found in our
observations. The importance of thermal instability is further supported by 
the large abundance of small HI clumps (see Section 4.2.1). 
The KH instability is important as well, however
it has a significantly longer timescale.
As we have shown, it is reasonable to expect that thermal and/or KH
fragments can survive for a long time, especially if surrounded by the ambient
gas with $T_{\rm h}<10^{6}$ K.

It is interesting that
cloud morphology in our observations is dominated by round shapes.
Elongated or head-tail clouds are surprisingly absent in our HI images.
Head-tail morphology is commonly observed for CHVCs and is often seen as a
signature of ram-pressure stripping, or other types of interactions between a
cloud and the ambient halo medium.
We therefore conclude that ram pressure effects
probably have a secondary role, with gravitational forces being 
the dominant gas removal process.

If we assume that thermal fragmentation is the dominant process responsible
for the clumpy structure, we can estimate a distance to the MS.
If we assume that the median value of the cloud angular 
size distribution (10$'$) corresponds to 
the linear size of `typical' thermal fragments, $\lambda_{\rm cool}\sim200$ pc 
(derived assuming $n_{\rm w}=5\times10^{-2}$ cm$^{-3}$), 
we estimate a distance of $\sim70$ kpc.
Please note that these numbers are only representative.
The cloud angular size varies and is also affected by our resolution limit, 
the volume density of the warm MS gas is also not a constant number. 
However, this exercise is very instructive and demonstrates that thermal
fragments can reasonably well explain at least some small-scale structure in the MS.   
The MS distance of $\sim70$ kpc is in agreement with the predictions from tidal models. 
Even more impressively, this distance agrees well with the 
recent estimate of 75 kpc by \cite{jin07}, who used  
a geometrodynamical model and assumed that
the MS follows a planar orbit around the Galactic center.


Also, the existence of multi-phase HI clouds can be used to place
an upper limit on the MS distance.
It is important to stress that the existence of multi-phase 
clouds along the tip of the MS, and especially at a distance 
of $\sim80$ kpc, is extremely surprising from a theoretical point of view.
Based on the consideration of cooling and heating processes
in an isothermal 10$^{6}$ K halo, \cite{Wolfire95} predicted 
that multi-phase clouds, pressure confined by the
hot halo, should exist in the MS only at distances less
than 20 kpc from the Galactic plane.
More recently, considering CHVCs \cite{Sternberg02}
showed that dark matter can provide an additional confinement mechanism, allowing
the existence of multi-phase structure at distances $\la150$ kpc.
The multi-phase clouds at the tip of the MS therefore suggest that 
the tip is likely to be at a distance $<150$ kpc. 
This argues against the existence of the third, very distant stream
found in the simulations by \cite{Connors06}, 
at least within the area covered by our observations.
Also, as pointed out by \cite{Kalberla06}, 
the multi-phase structure at the MS distance suggests a
lower confining halo gas pressure than what is expected in 
the theoretical framework.

\section{Summary and Conclusions}

We have obtained HI observations of a region, $\sim900$ square degrees large, 
at the tip of the MS. The major results are as follows.

\begin{enumerate}
\item We find four coherent HI streams at the tip of the MS, extending
  continuously from Dec 5\degree~to $25$\degree. 
The streams have different morphology and velocity gradients. 
Three streams are very clumpy, while one stream is mainly made up
of large diffuse structures.

\item The comparison of observations with the tidal simulations
by \cite{Connors06} suggests that the three clumpy streams could be a result
of a 3-way tidal splitting of the main MS filament, induced by encounters
between the LMC and the MS. 
The fourth stream is probably much younger, and could have been tidally drawn from the
Magellanic Bridge in the more recent past, $<200-500$ Myr. This stream is more diffuse
because it has not had enough time to cool and fragment.

\item We find an extensive population of HI clouds at the tip of the MS. The
smallest clouds have a size reaching down to our resolution limit of 3.5$'$.
70\% of clouds have an angular size in the range 3.5$'$--10$'$.
About 12\% of the MS clouds show multi-phase velocity profiles, 
indicating the co-existence of cooler and warmer gas. 
Cloud central velocities show two peaks at $-350$ and $-400$ \kms.

\item We show that for warm gas trailing through the Galactic halo for the
past Gyr or so, the effects of KH instability are relatively unimportant. 
However, thermal instability, with a typical timescale of 
a few Myrs to a few hundred Myrs, is significantly important and will result
in the fragmentation of gas and the formation of smaller clumps. The 
observed clumpiness at the tip of the MS  could be fully or partially due to
this mechanism.

\item We show that thermal/KH fragments can survive in the hot halo for a long time ($>1$ Gyr),
especially if surrounded by a $<10^{6}$ K halo gas.

\item If the observed angular size of HI clumps is due to thermal instability, then
the tip of the MS is at a distance of $\sim70$ kpc.

\end{enumerate}

\acknowledgements
We are grateful to everyone at the Arecibo Observatory for
their help in conducting these observations.
It is a pleasure to acknowledge
stimulating discussions with Fabian Heitsch, Bruce Elmegreen, 
Bob Benjamin, and Jay Gallagher. 
We thank an anonymous referee for constructive suggestions.
Support by NSF grants  AST-0097417 and AST-0707679 is gratefully acknowledged.


\begin{thebibliography}{34}
\expandafter\ifx\csname natexlab\endcsname\relax\def\natexlab#1{#1}\fi

\bibitem[{{Arnal} {et~al.}(2000){Arnal}, {Bajaja}, {Larrarte}, {Morras}, \&
  {P{\"o}ppel}}]{Arnal00}
{Arnal}, E.~M., {Bajaja}, E., {Larrarte}, J.~J., {Morras}, R., \& {P{\"o}ppel},
  W.~G.~L. 2000, \aaps, 142, 35

\bibitem[{Audit \& Hennebelle(2005)}]{Audit05}
Audit, E. \& Hennebelle, P. 2005, A\&A, 433, 1

\bibitem[{{Bajaja} {et~al.}(2005){Bajaja}, {Arnal}, {Larrarte}, {Morras},
  {P{\"o}ppel}, \& {Kalberla}}]{Bajaja05}
{Bajaja}, E., {Arnal}, E.~M., {Larrarte}, J.~J., {Morras}, R., {P{\"o}ppel},
  W.~G.~L., \& {Kalberla}, P.~M.~W. 2005, \aap, 440, 767

\bibitem[{{Besla} {et~al.}(2007){Besla}, {Kallivayalil}, {Hernquist},
  {Robertson}, {Cox}, {van der Marel}, \& {Alcock}}]{Besla07}
{Besla}, G., {Kallivayalil}, N., {Hernquist}, L., {Robertson}, B., {Cox},
  T.~J., {van der Marel}, R.~P., \& {Alcock}, C. 2007, ArXiv Astrophysics
  e-prints

\bibitem[{{Braun} \& {Thilker}(2004)}]{Braun04b}
{Braun}, R. \& {Thilker}, D.~A. 2004, \aap, 417, 421

\bibitem[{{Br{\"u}ns} {et~al.}(2005){Br{\"u}ns}, {Kerp}, {Staveley-Smith},
  {Mebold}, {Putman}, {Haynes}, {Kalberla}, {Muller}, \& {Filipovic}}]{Bruns05}
{Br{\"u}ns}, C., {Kerp}, J., {Staveley-Smith}, L., {Mebold}, U., {Putman},
  M.~E., {Haynes}, R.~F., {Kalberla}, P.~M.~W., {Muller}, E., \& {Filipovic},
  M.~D. 2005, \aap, 432, 45

\bibitem[{{Connors} {et~al.}(2006){Connors}, {Kawata}, \& {Gibson}}]{Connors06}
{Connors}, T.~W., {Kawata}, D., \& {Gibson}, B.~K. 2006, \mnras, 371, 108

\bibitem[{{de Heij} {et~al.}(2002){de Heij}, {Braun}, \& {Burton}}]{deHeij02}
{de Heij}, V., {Braun}, R., \& {Burton}, W.~B. 2002, \aap, 391, 67

\bibitem[{Gardiner \& Noguchi(1996)}]{Gardiner96}
Gardiner, L.~T. \& Noguchi, M. 1996, MNRAS, 278, 191

\bibitem[{{Giovanelli} {et~al.}(2005){Giovanelli}, {Haynes}, {Kent},
  {Perillat}, {Saintonge}, {Brosch}, {Catinella}, {Hoffman}, {Stierwalt},
  {Spekkens}, {Lerner}, {Masters}, {Momjian}, {Rosenberg}, {Springob},
  {Boselli}, {Charmandaris}, {Darling}, {Davies}, {Lambas}, {Gavazzi},
  {Giovanardi}, {Hardy}, {Hunt}, {Iovino}, {Karachentsev}, {Karachentseva},
  {Koopmann}, {Marinoni}, {Minchin}, {Muller}, {Putman}, {Pantoja}, {Salzer},
  {Scodeggio}, {Skillman}, {Solanes}, {Valotto}, {van Driel}, \& {van
  Zee}}]{Giovanelli05}
{Giovanelli}, R., {Haynes}, M.~P., {Kent}, B.~R., {Perillat}, P., {Saintonge},
  A., {Brosch}, N., {Catinella}, B., {Hoffman}, G.~L., {Stierwalt}, S.,
  {Spekkens}, K., {Lerner}, M.~S., {Masters}, K.~L., {Momjian}, E.,
  {Rosenberg}, J.~L., {Springob}, C.~M., {Boselli}, A., {Charmandaris}, V.,
  {Darling}, J.~K., {Davies}, J., {Lambas}, D.~G., {Gavazzi}, G., {Giovanardi},
  C., {Hardy}, E., {Hunt}, L.~K., {Iovino}, A., {Karachentsev}, I.~D.,
  {Karachentseva}, V.~E., {Koopmann}, R.~A., {Marinoni}, C., {Minchin}, R.,
  {Muller}, E., {Putman}, M., {Pantoja}, C., {Salzer}, J.~J., {Scodeggio}, M.,
  {Skillman}, E., {Solanes}, J.~M., {Valotto}, C., {van Driel}, W., \& {van
  Zee}, L. 2005, \aj, 130, 2598

\bibitem[{{Hartmann} \& {Burton}(1997)}]{Hartmann97}
{Hartmann}, D. \& {Burton}, W.~B. 1997, {Atlas of Galactic Neutral Hydrogen}
  (Atlas of Galactic Neutral Hydrogen, by Dap Hartmann and W.~Butler Burton,
  pp.~243.~ISBN 0521471117.~Cambridge, UK: Cambridge University Press, February
  1997.)

\bibitem[{{Heiles}(2007)}]{Heiles07}
{Heiles}, C. 2007, \pasp, 119, 643


\bibitem[{{Hoffman} {et~al.}(2004){Hoffman}, {Salpeter}, \&
  {Hirani}}]{Hoffman04}
{Hoffman}, G.~L., {Salpeter}, E.~E., \& {Hirani}, A. 2004, \aj, 128, 2932

\bibitem[{Jin \& Lynden-Bell(2007)}]{jin07}
Jin, S. \& Lynden-Bell, D. 2007, MNRAS

\bibitem[{{Kalberla} {et~al.}(2005){Kalberla}, {Burton}, {Hartmann}, {Arnal},
  {Bajaja}, {Morras}, \& {P{\"o}ppel}}]{Kalberla05a}
{Kalberla}, P.~M.~W., {Burton}, W.~B., {Hartmann}, D., {Arnal}, E.~M.,
  {Bajaja}, E., {Morras}, R., \& {P{\"o}ppel}, W.~G.~L. 2005, \aap, 440, 775

\bibitem[{{Kalberla} \& {Haud}(2006)}]{Kalberla06}
{Kalberla}, P.~M.~W. \& {Haud}, U. 2006, \aap, 455, 481

\bibitem[{{Kallivayalil} {et~al.}(2006){Kallivayalil}, {van der Marel}, \&
  {Alcock}}]{Kallivayalil06}
{Kallivayalil}, N., {van der Marel}, R.~P., \& {Alcock}, C. 2006, \apj, 652,
  1213

\bibitem[{{Lin} \& {Murray}(2000)}]{Lin00}
{Lin}, D.~N.~C. \& {Murray}, S.~D. 2000, \apj, 540, 170

\bibitem[{{Mastropietro} {et~al.}(2005){Mastropietro}, {Moore}, {Mayer},
  {Wadsley}, \& {Stadel}}]{Mastropietro05}
{Mastropietro}, C., {Moore}, B., {Mayer}, L., {Wadsley}, J., \& {Stadel}, J.
  2005, \mnras, 363, 509

\bibitem[{Mathewson {et~al.}(1974)Mathewson, Cleary, \& Murray}]{Mathewson74}
Mathewson, D.~S., Cleary, M.~N., \& Murray, J.~D. 1974, ApJ, 190, 291

\bibitem[{{McKee} \& {Cowie}(1977)}]{McKee77b}
{McKee}, C.~F. \& {Cowie}, L.~L. 1977, \apj, 215, 213

\bibitem[{{Murray} {et~al.}(1993){Murray}, {White}, {Blondin}, \&
  {Lin}}]{Murray93}
{Murray}, S.~D., {White}, S.~D.~M., {Blondin}, J.~M., \& {Lin}, D.~N.~C. 1993,
  \apj, 407, 588

\bibitem[{{Nidever} {et~al.}(2007){Nidever}, {Majewski}, \& {Butler
  Burton}}]{Nidever07}
{Nidever}, D.~L., {Majewski}, S.~R., \& {Butler Burton}, W. 2007, ArXiv
  e-prints, 706

\bibitem[{{Putman} {et~al.}(2003){Putman}, {Staveley-Smith}, {Freeman},
  {Gibson}, \& {Barnes}}]{Putman03}
{Putman}, M.~E., {Staveley-Smith}, L., {Freeman}, K.~C., {Gibson}, B.~K., \&
  {Barnes}, D.~G. 2003, \apj, 586, 170

\bibitem[{{Sembach} {et~al.}(2003){Sembach}, {Wakker}, {Savage}, {Richter},
  {Meade}, {Shull}, {Jenkins}, {Sonneborn}, \& {Moos}}]{Sembach03}
{Sembach}, K.~R., {Wakker}, B.~P., {Savage}, B.~D., {Richter}, P., {Meade}, M.,
  {Shull}, J.~M., {Jenkins}, E.~B., {Sonneborn}, G., \& {Moos}, H.~W. 2003,
  \apjs, 146, 165

\bibitem[{{Slavin}(2007)}]{Slavin07}
{Slavin}, J.~D. 2007, in Astronomical Society of the Pacific Conference Series,
  Vol. 365, SINS - Small Ionized and Neutral Structures in the Diffuse
  Interstellar Medium, ed. M.~{Haverkorn} \& W.~M. {Goss}, 113



\bibitem[{{Stanimirovi{\' c}} {et~al.}(2002){Stanimirovi{\' c}}, {Dickey},
  {Kr{\v c}o}, \& {Brooks}}]{Stanimirovic02}
{Stanimirovi{\' c}}, S., {Dickey}, J.~M., {Kr{\v c}o}, M., \& {Brooks}, A.~M.
  2002, \apj, 576, 773

\bibitem[{{Stanimirovi{\'c}} {et~al.}(2006){Stanimirovi{\'c}}, {Putman},
  {Heiles}, {Peek}, {Goldsmith}, {Koo}, {Kr{\v c}o}, {Lee}, {Mock}, {Muller},
  {Pandian}, {Parsons}, {Tang}, \& {Werthimer}}]{Stanimirovic06}
{Stanimirovi{\'c}}, S., {Putman}, M., {Heiles}, C., {Peek}, J.~E.~G.,
  {Goldsmith}, P.~F., {Koo}, B.-C., {Kr{\v c}o}, M., {Lee}, J.-J., {Mock}, J.,
  {Muller}, E., {Pandian}, J.~D., {Parsons}, A., {Tang}, Y., \& {Werthimer}, D.
  2006, \apj, 653, 1210

\bibitem[{Stanimirovi\'{c} {et~al.}(1999)Stanimirovi\'{c}, Staveley-Smith,
  Dickey, Sault, \& Snowden}]{Stanimirovic99}
Stanimirovi\'{c}, S., Staveley-Smith, L., Dickey, J.~M., Sault, R.~J., \&
  Snowden, S.~L. 1999, MNRAS, 302, 417

\bibitem[{{Sternberg} {et~al.}(2002){Sternberg}, {McKee}, \&
  {Wolfire}}]{Sternberg02}
{Sternberg}, A., {McKee}, C.~F., \& {Wolfire}, M.~G. 2002, \apjs, 143, 419

\bibitem[{{Wakker} \& {van Woerden}(1991)}]{Wakker91}
{Wakker}, B.~P. \& {van Woerden}, H. 1991, \aap, 250, 509

\bibitem[{Wannier \& Wrixon(1972)}]{Wannier72}
Wannier, P. \& Wrixon, G.~T. 1972, \apjl, 173, L119

\bibitem[{Wolfire {et~al.}(1995)Wolfire, McKee, Hollenbach, \&
  Tielens}]{Wolfire95}
Wolfire, M.~G., McKee, C.~F., Hollenbach, D., \& Tielens, A. G. G.~M. 1995,
  \apj, 453, 673

\bibitem[{Yoshizawa \& Noguchi(1999)}]{Yoshizawa99}
Yoshizawa, A.~M. \& Noguchi, M. 1999, in IAU Symp. 186: Galaxy Interactions at
  Low and High Redshift, Vol. 186, 60

\end{thebibliography}

\label{lastpage}
\end{document}